\documentclass%
[prl,amsmath,showpacs,twocolumn,preprintnumbers,superscriptaddress]%
{revtex4}

\usepackage{mathptmx}

\usepackage{microtype}

\usepackage{graphicx}
\usepackage{dcolumn}
\usepackage{amsmath}
\usepackage{mathtools}
\usepackage{bm}

\usepackage{mathptmx}
\usepackage{microtype}

\usepackage{graphicx}
\usepackage{dcolumn}
\usepackage{amsmath}
\usepackage{mathtools}
\usepackage{bm}
\usepackage{amssymb}

\newcommand*{\w}{\omega}

\DeclarePairedDelimiter{\abs}{\lvert}{\rvert}
\newcommand*{\Const}{\text{\textit{Const}}}

\begin{document}

\title{Majorana states for subluminal structured photons}

\author{Fabrizio Tamburini} 
\email{fabrizio.tamburini@gmail.com}
\affiliation{ZKM -- Zentrum f\"ur Kunst und Medientechnologie, Lorentzstr. 19, D-76135, Karlsruhe, Germany.}
\affiliation{MSC -- BW,  Stuttgart, Nobelstr. 19, D-70569 Stuttgart, Germany.}

\author{Bo Thid\'e}
\email{bt@irfu.se}
\affiliation{Swedish Institute of Space Physics,
 {\AA}ngstr\"{o}m Laboratory, P.\,O.~Box~537, SE-75121, Sweden} 
 
\author{Ignazio Licata}
\affiliation{Institute for Scientific Methodology (ISEM) Palermo Italy; ignazio.licata@ejtp.info}
\affiliation{School of Advanced International Studies on Theoretical and Nonlinear Methodologies of Physics, Bari, I-70124, Italy}
\affiliation{International Institute for Applicable Mathematics and Information Sciences (IIAMIS), B.M. Birla Science Centre, Adarsh Nagar, Hyderabad -- 500 463, India}

\author{Fr\'ed\'eric Bouchard}
\affiliation{Department of Physics, University of Ottawa, 25 Templeton St., Ottawa, Ontario, K1N 6N5 Canada}

\author{Ebrahim Karimi}
\affiliation{Department of Physics, University of Ottawa, 25 Templeton St., Ottawa, Ontario, K1N 6N5 Canada}
\affiliation{Department of Physics, Institute for Advanced Studies in Basic Sciences, 45137-66731 Zanjan, Iran}

\begin{abstract}

The speed of light in vacuum, $c$, is a fundamental constant of nature. Photons belonging to a structured beam of finite transverse size, generated by a spatial light modulator, have been observed to travel with a group velocity, $v_g$, which is smaller than $c$ also when propagating in vacuum~\cite{padgett,karimi,bareza}. This is an effect that depends on the geometry of the beam. From quantum mechanical considerations, these photons must in any case propagate at the speed of light. This paradox can be resolved by taking into account a projection effect. What was measured in these experiments as group velocity was in fact its projection onto the beam propagation axis~\cite{horvath}. This depends on the divergence of the beams used in these experiments. We have found that for hypergeometric beams carrying orbital angular momentum (OAM), generated by sources with equal aperture~\cite{karimi:07,kummer,kummer2,anzo}, $v_g$ obeys an OAM/velocity relationship similar to that proposed by Majorana between spin and mass for bosonic and fermionic relativistic particles. This relationship, depending on the geometrical properties of the beam, can pave the way for an alternative estimation of OAM or to implement a time buffer in optical communications.
\end {abstract}


\maketitle

\section{Introduction}

Photons in structured light beams and with a finite transverse size were observed to travel at a speed apparently slower than that of light, including in vacuum~\cite{padgett,karimi,bareza}. Photons propagating in vacuum, however, must still propagate at the speed of light; this behaviour was explained as a projection effect of the effective motion of the photon on the axis of propagation of the diverging beam, thus depending on the geometrical properties of the beam \cite{horvath}.  In this scenario, we have found that structured beams carrying OAM, and more specifically, hypergeometric beams, present a group velocity $v_g$ that obeys a relationship similar to that proposed by Majorana between spin and mass for bosonic and fermionic relativistic particles that instead involves OAM, speed of propagation and a virtual mass parameter that characterizes the beam. The mathematical structure of the Majorana solution is also known as the ``Majorana Tower''~\cite{tower2}.

Majorana formulated in 1932, and then again in 1937, an alternative solution valid for bosonic and fermionic relativistic particles with null or positive-definite rest mass in the attempt to avoid the problem of the negative squared mass solution emerging from the Dirac equation~\cite{Majorana:NC:1932,majorana1937teoria} that instead were due to anti-electrons, experimentally discovered by Anderson~\cite{PhysRev.43.491}. In a generalization of the Dirac equation to spin values different than that of the electron, Majorana found a solution with a denumerable infinite spectrum of particles that obey either the Bose-Einstein or the Fermi-Dirac statistics with a precise relationship between spin and mass. The infinite spin solution to the Dirac equation proposed by Majorana gives a spectrum of particles with a positive-definite or null finite squared mass. The spectrum of particles described by the Majorana solution, the ``Majorana Tower''~\cite{tower2}, does not have any correspondence with that of the Standard Model~\cite{2002PhRvD..66a0001H}. In this scenario, particles and their corresponding antiparticles are not mutually distinguishable. Since any particle and antiparticle must have opposite electric charge, this type of solution is valid only for known neutral particles. This relationship applies only to bosons such as gravitons and photons, that are zero rest-mass particles in vacuum and to a few possible exceptions for a particular class of fermions, the so-called Majorana neutrinos~\cite{Doi:1985dx,Mohapatra:2005wg,Avignone:2007fu}. Clearly, elementary particles such as electrons and positrons cannot be Majorana particles.

As known from the study of Majorana's unpublished works~\cite{esposito}, Ettore Majorana progressively came to the idea of the relativistic theory of particles with arbitrary angular momentum (``\emph{Teoria Relativistica di Particelle Con Momento Intrinseco Arbitrario}", 1932). He studied first the finite cases for composite systems, and in particular a Dirac-like equation for photons, where he faced the problem by starting directly from the Maxwell classical field, using the 3-dimensional complex vector $F = E + iH$, where $E$ and $H$ are the electric and magnetic field in electrostatic units (esu), respectively~\cite{mignani}. This expression sets two invariants for the electromagnetic tensor, one for the real and one for the imaginary part of F, and introduces a \textit{wavefunction} of the photon $\Psi  = E  \pm i H$, whose probabilistic interpretation is very simple and immediate and relies directly the initial intuition of Einstein-Born: the electromagnetic energy density is proportional to the probability density of the photons. Maxwell's equations can be written in terms of this wavefunction; the first is the typical transverse state in quantum mechanics for the spin-1 particles, while developing the second Majorana is capable of writing the photon wave equation as a particular case of a Dirac equation, which is shown to be equivalent to quantum electrodynamics (QED)~\cite{tamvicino,iwo:96}. 

This was the starting point for Ettore Majorana's ``theory of everything". It is important to note that the theory describes a single particle at rest and fields with variable spin, i.e., not elementary but composite systems, and applies to both bosons and fermions, anticipating the latest supermultiplet of supersymmetric (SUSY) theories. In other words, the work by Majorana consists of having extracted the most general conditions of space-temporal symmetry for a Hilbert Space from a great mathematical phenomenology where all the spins can be represented by the non-homogeneous Lorentz group. This space is infinite dimensional and can be tied to a single spin with appropriate contour conditions. Therefore, Majorana's theory is ideal for studying bound systems such as almost particle or photon systems in a structured beam, such as those we look at in this work. For theoretical developments of the theory see also refs.~\cite{gelfand,valramov}.

Examples of Majorana-like ``particles'' can be found in different scenarios like in condensed matter physics~\cite{wilczek}, where composite states of particles behave like Majorana fermions~\cite{ferrara15}, for which a particle and its antiparticle must coincide and acquire mass from a self-interaction mechanism. This is different from the Standard Model which obeys the Higgs mechanism~\cite{higgs}. These quasiparticles are the product of electromagnetic interactions between electrons and atomic structures present in a condensed matter scenario.

It is at all effects an emergent behaviour that follows the mathematical rules put down for Majorana fermions. Quasiparticle excitations behaving like Majorana particles have been observed in topological superconductors~\cite{fukane08,neupert} characterized by carrying null electric charge and energy~\cite{kawa15}. Other types of quasiparticles, called Majorana zero modes, were observed in Josephson junctions~\cite{ohm14} and in solid state systems. They hold promise for interesting future applications in information processing~\cite{sarma15,borstenduff,kitaev} and photonic systems~\cite{mzms}. For a deeper insight see ref.~\cite{elliott2015}.

In addition, structured light beams and other photonic systems can behave like Majorana particles. They involve not only the spin angular momentum but also the total angular momentum. Photons, in fact, carry energy, momentum and angular momentum $\mathbf{J}$.  The conserved angular momentum quantity carried by a photon is given by the (vector) sum of the spin angular momentum (SAM), $\boldsymbol{\Sigma}$, and the orbital angular momentum (OAM),~$\mathbf{L}$.

An example is photons carrying OAM propagating in a structured plasma~\cite{Tamburini&al:EPL:2010,tower}. They acquire an effective Proca mass through the Anderson-Higgs mechanism~\cite{Anderson:PR:1963,higgs} and obey a mass/OAM relationship that resembles the mathematical structure of the Majorana tower.  The difference between the original solution found by Majorana, based on the space-time symmetries of the Lorentz group applied to the Dirac equation in vacuum, and that of photons in a structured plasma, is the Anderson-Higgs mechanism and the role played by OAM for the mass/total angular momentum relationship found for structured electromagnetic beams.

Similar to what occurs in Majorana's original spin/mass relationship, OAM acts as a term that reduces the total Proca photon mass in the plasma~\cite{Tamburini&al:EPL:2010,tower}. What actually induces a Proca mass in photons in a plasma is the breaking of spatial homogeneity and a characteristic scale length introduced by the plasma structure at the frequencies at which the plasma is resonant. The presence of a characteristic scale length and structures in the plasma presents strong analogies with the models of space-time characterized by a modified action of the Lorentz group, such as the Magueijo-Smolin model~\cite{PhysRevLett.88.190403}, demonstrating deep analogies with the dynamics described by the Dirac equation when a characteristic scale length is present. Unavoidably, Berry phase effects are introduced~\cite{Gosselin&al:PLB:2008}. Fermionic and anionic-like behaviours are
obtained through the coherent superposition of OAM states.

It is of great interest to note that the effective Proca mass falls under the general conditions provided by the Higgs mechanism. When a local gauge symmetry breaks up, Goldstone bosons (no massive scalar particles) may be missing or some gauge fields become massive. In the Abelian Higgs model, a mass term for the spin field 1 (just like in Proca's action) emerges and a mass term for the real Higgs scalar field. It is possible to show that Maxwell-Proca equations fall into the affine-Higgs case studied by Stueckelberg, a powerful covariance principle between free space wavelength equations and those in a conservative force field~\cite{stueckelberg}. Today there is a renewed interest in Stueckelberg models for non-abelian gauge theories in the possibility of a re-reading of the Higgs mechanism \cite{ruegg,aldaya,nisino}.

\section{Majorana tower for OAM standard paraxial solutions}
The apparent slowing down of light in a structured beam in vacuum finds a geometrical meaning related to the beam divergence. Hence, it is slightly different from the mechanism in which light is slowed due to the presence of matter. Structured beam photons propagating in vacuum may show a different behaviour than that of a plane wave because of the field confinement induced by the finite extent and the structure of the beam that changes the wavevector with the result of altering the group velocity, $v_g$, measured along the propagation axis.

What is of fundamental importance is that also the phase velocity $v_p$ is different for different OAM modes. The heart of the matter regarding the speed of propagation of OAM eigenmodes, propagating along the $z$~axis, lies in the fact that $k_z$ is not the only component of the wave vector that has to be taken into account in order to have a complete description of the system. If we consider a distinct $z$ component of OAM, $J_z$, as a single eigenmode, characterized by its azimuthal quantum number $m$, where~$\ell$ can take any integer value ranging from $-\abs{\ell}$ to $+\abs{\ell}$, we also have a $\phi$ component of the $k$ vector that varies in time since we have a rotational degree of freedom. This additional component has implications on how to interpret the concept of phase velocity. Moreover, also the radial component of the wavevector, $k_\rho$, related to the $p$ quantum number must be taken into account. Thus, the effective magnitude of $k$ is no longer $k_z$ in the direction of propagation of the symmetry axis of the beam, but the wavevector that identifies the propagation of photons is $\sqrt{(k_\rho)^2+ (\ell/\rho)^2 + (k_z)^2}$.

The consequence of this result is that, for a given frequency $\w$, one finds a different phase velocity $v_p$ for each OAM quantum number $\ell$ when the $k$~vector no longer points along the $z$-axis. Instead it varies in time, precess along a cone with its axis along the $z$-axis.


To better explain this concept, consider a general electric field in cylindric coordinates $\mathbf{E}(\rho,\phi,z;t)$. In the most general case, when the field is factorable
\begin{equation}
	\mathbf{E}(\rho,\phi,z;t)=R(\rho) \Phi(\phi)Z(z)T(t) \mathbf{E}_0,
\end{equation}
and $\mathbf{E}_0=E_0 \mathbf{\hat e}$; the field is propagating along $z$. The azimuthal function $\Phi(\phi)$ that describes the variation of \textbf{E} in the direction perpendicular to both $z$ and $\rho$ can most conveniently be expanded in a Fourier series with constant amplitude coefficients $c_m$
\begin{equation}
	\Phi(\phi)=\sum^{+\infty}_{\ell=-\infty} c_\ell\,e^{i \ell \phi}
\end{equation}
that leads to an expansion of the electric field in discrete angular momentum modes $\mathbf{E}_\ell\propto e^{i \ell \phi}$. The azimuthal component of the wavevector is $k_\phi=\ell/\rho$ and the wavevector, expanded along the coordinates, does not coincide with the direction of propagation along the $z$-axis. Instead it becomes
\begin{equation}
	k_\ell(\rho_0) = k_\mathbf{\rho} \mathbf{\hat \rho} + \frac{\ell}{\rho_0} \mathbf{\hat \phi} + k_z \hat z,
\end{equation}
where $\rho_0$ represents the radial mode. From this, one obtains a phase velocity $v_p$ in vacuum
\begin{equation}
	v_p = \frac{\w}{k}= \frac{\w}{k_z\sqrt{1+\left(\frac{k_\rho}{k_z}\right)^2+\left(\frac{\ell}{k_z\rho_0}\right)^2}}.
\end{equation}
Since $\abs{k_z} \leq \abs{k}$, the phase velocity projected onto the $z$~axis will always appear superluminal
\begin{equation}
	v_{p,z} = \frac{\w}{k_z} \geq c .
\end{equation}
For a plane wave propagating in vacuum, it is conventionally assumed that the magnitude of the wave vector $k$ is identical to $k_z$ (the component of the $k$ vector along the $z$~axis, conveniently and customarily set to be the line of sight from the source to the observer.  For the group velocity, $v_g$, the concept is the same, without assuming \emph{a priori} constraints on the geometry of the beam~\cite{reviewpaper}.

A different case is a Bessel beam where the scalar amplitude is $J_\ell(k_\rho \rho)e^{ik_z z}e^{i \ell\phi}e^{-i\omega t}$, which carries OAM with $\ell\hbar$ angular momentum per photon. There will be an azimuthal component of Poynting's vector but there is no corresponding $\phi$ component of the wavevector. To see this we note that $\omega^2 = (k_z^2 + k_\rho^2)c^2$. In this case, there is not azimuthal component of ${\bf k}$.

Because of the relationship $v_p v_g = c^2$, the group velocity $v_{g,z}$, projected onto and measured along the $z$-axis, appears to be subluminal as reported in refs.~\cite{padgett,bareza,horvath}. This explains the paradoxical behaviour of photons that belong to particular classes of structured beams seem to propagate in vacuum with an apparent group velocity $v_{g,z}<c$ without violating the fundamental laws of relativity or quantum mechanics.  This is due to the projection effect that depends on the peculiar geometric properties of the class of OAM beams considered.  For the sake of simplicity in the notation, from here on, we will consider the projections of the phase and group velocities along the propagation axis (the $z$-axis).

We will now focus our attention on the class of beams obtained by imposing OAM onto beams of finite widths through a spiral phase plate or a fork hologram~\cite{oambook}. These beams---widely used in standard OAM experiments---are described in terms of Hypergeometric-Kummer functions~\cite{karimi:07,kummer,kummer2,anzo} and belong to the subclass of circular beams~\cite{pino} for which a precise linear relationships between their divergence and the OAM eigenvalue $\ell$ is present \cite{anzo,oldoni1,oldoni2}.  In this case, $v_g$ is reduced by an amount that depends upon the aperture of the optical system; the
time delay between different beams caused by different values of the group velocity was proposed to set up a temporal buffer for optical computing and information transfer in free space~\cite{alfano16,coles17}.

Consider this general subclass of OAM beams emitted by one and the same finite-sized aperture. A transverse spatial confinement of the field leads to a larger modification of the axial component of the wavevector $k_z$, modifying the group velocity $v_g$, when $\abs{\ell}$ increases. For an OAM beam propagating along the $z$-axis, the wavevector is ${k_0=2\pi/\lambda}$ and ${k^2_0=k^2_x+k^2_y+k^2_z}$, and the confined mode is given by the quantities $k_x,~k_y>0$ that imply $k_z<k_0$ with a modification in both the phase and group velocities, $v_p$ and $v_g$, respectively.

When the beam is confined, as occurs in the class of Hypergeometric-Kummer beams, the relationship $k_x,k_y=\Const$ implies that $k_z$ is dispersive in free space, similar to what happens in the case of Bessel beams. In radial coordinates, this relationship becomes $k_r=\sqrt{k^2_x+k^2_y}$ and ${k_z=k_0-(k^2_\rho/2k^2_0)}$.

For these particular types of OAM-carrying beams, the phase velocity $v_p$, measured along the $z$ direction, is
\begin{equation}
	v_p=\frac c{\left(1-\frac{k^2_\rho}{2 k^2_0}\right)},
\label{fase}
\end{equation}
and the group velocity along $z$ is
\begin{equation}
	v_{g}=c\left(1-\frac{k^2_\rho}{2 k^2_0}\right).
\label{gruppo}
\end{equation}
Hence, the group velocity reduction $v_g<c$ gives a measurable delay that reveals the OAM content of the beam~\cite{bareza,bareza2,saari,alfano17}.
%
%
The group velocity can then be expressed in terms of the transverse quantum numbers $p$ and $\ell$~\cite{bareza},
\begin{equation}
	v_g=\frac{c}{1+\left(\frac{\theta_0}{4}\right)^2 (2p+|\ell |+ 1) }
\end{equation}
In fact, from this delay, the dispersion in free-space of OAM beams can be characterized by a far-field beam divergence angle $\theta_0$, which is related to the radial and azimuthal indices $p$ and $\ell$, respectively. As discussed in ref.~\cite{bareza}, the group velocity for Laguerre-Gauss beams, for example, depends on the divergence angle, and consequently on the OAM value of the beam.

\begin{figure}[!htbp]
	\begin{center}
	\includegraphics[width=1\columnwidth]{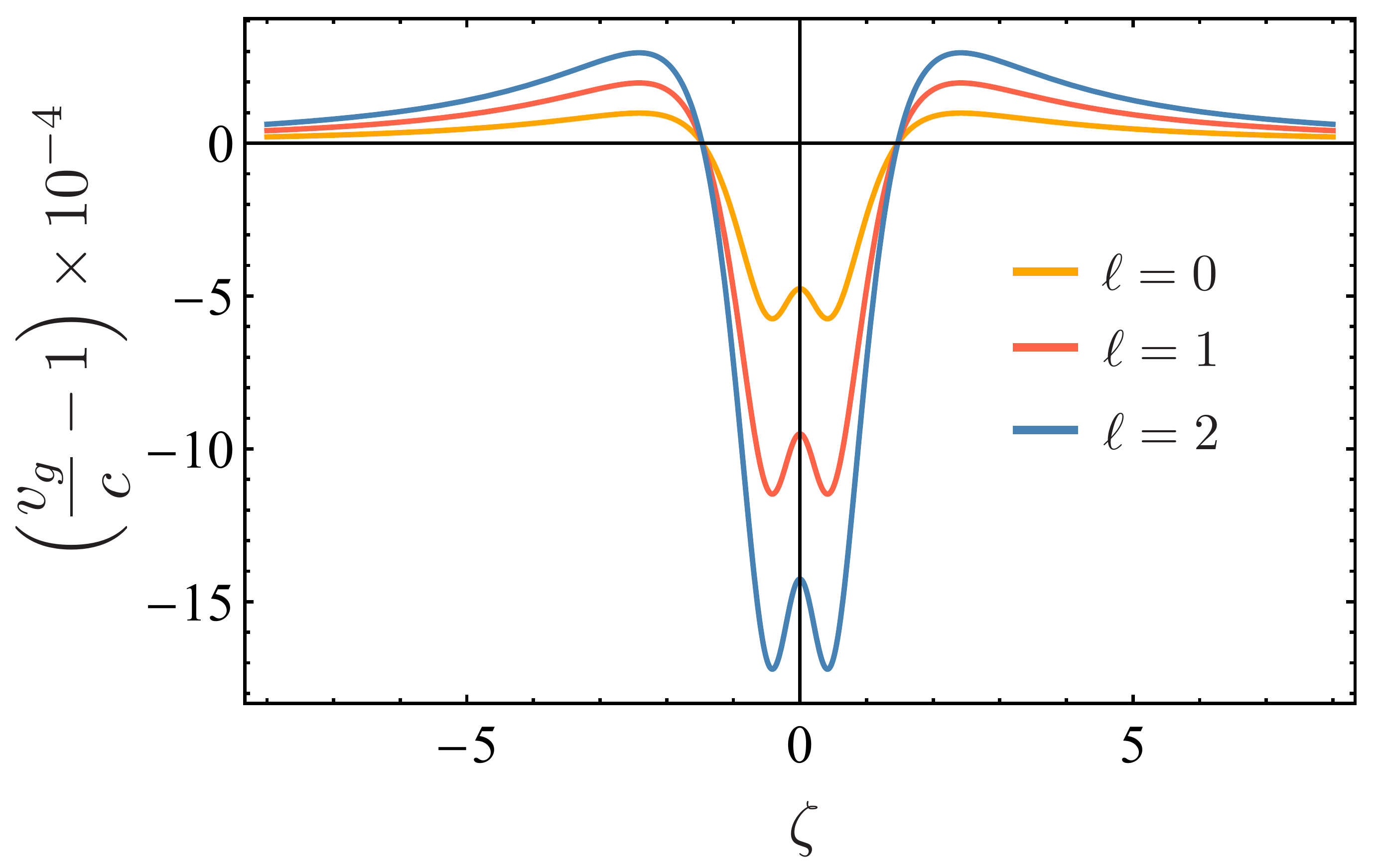}
	\caption[]{Group velocity as a function of the dimensionless propagation distance $\zeta$ for various $\ell$ values. The divergence angle is set to $\theta_0=\pi/36$.}
	\label{fig1}
	\end{center}
\end{figure}
Furthermore, it is possible to calculate the exact group velocity of a paraxial beam at any point in space using the \textit{wave picture} description. In the wave description, the group velocity is given by $v_g=|\partial_\omega \nabla \Phi |^{-1}$, where $\Phi(\mathbf r)$ represents the wave phase front. For Laguerre-Gauss modes, the group velocity along $z$, which has an explicit dependence on the propagation distance $z$, is given by 
\begin{equation}
	v_g=\frac{c}{1+\left(\frac{\theta_0}{4}\right)^2 (2p+|\ell |+ 1) {\cal F}(\zeta)},
\end{equation}
where
\begin{align}
	{\cal F}(\zeta)=\frac{\left(1+6 \zeta^2 - 3 \zeta^4 \right)}{\left( 1+\zeta^2 \right)^3},
\end{align}
$\zeta:=z/z_\mathrm{R}$ is a dimensionless coordinate and $z_\mathrm{R}=8/(k \theta_0^2)$ is the Rayleigh range. The full description of the group velocity gives rise to a subluminal behaviour, for $|\zeta|<\sqrt{1+2 \sqrt{3}/3}$ and a superluminal behaviour for $|\zeta|>\sqrt{1+2 \sqrt{3}/3}$. The subluminal and superluminal behaviour of the Laguerre-Gauss modes upon propagation are shown in Fig.~\ref{fig1} for beams carrying OAM values of $\ell=1,2$ and $3$.


This relationship reflects the peculiar geometry of these beams. The time delays--or, better, different group velocities--can be interpreted as the effect of a fictitious mass $m_v$ of a quasiparticle state that characterizes the dynamics of the beam and is described by a Schr\"odinger-like equation at low velocities $v_g$, propagating with the group velocity $v_g$,
\begin{equation}
 v_g= \frac 1\hbar \nabla_{k_z} \frac{\hbar^2 k_z^2}{2 m_v}
  = \frac{\hbar k_z}{m_v}
  = \frac p m_v .
\end{equation}
%
Any of these quasiparticle states have a precise relationship between their angular momentum value and their fictitious mass in vacuum that resembles the behaviour of Proca photons with OAM in a plasma.  

This means that the process of beam confinement and structuring plays a role similar to that of the Anderson-Higgs mechanism that induces the Proca mass on the photons together with their angular momentum/mass relationship cast in a Majorana tower, similar to what occurs in waveguides even if the beams are freely propagating in space.

At all effects, photons propagating in vacuum belonging to this class of structured beams obey the mathematical rules of a Majorana tower of quasiparticle states.  In fact, by comparing the two group velocities one obtains a relationship that involves the wavevector $k$ and the fictitious mass term $m_v$
\begin{equation}
	\frac{\hbar k_z}{m_v}= \frac{c}{1+\left(\frac{\theta_0}{4}\right)^2 (2p+\abs{\ell}+ 1){\cal F}(\zeta)}
\end{equation}
from this one obtains the fictitious mass value for this quasiparticle state, that reflects the geometrical properties of these beams,
\begin{equation}
	m_v =\frac{\hbar k_z}{c} \left[ 1+\left(\frac{\theta_0}{4}\right)^2(2p+\abs{\ell}+1){\cal F}(\zeta) \right].
\end{equation}

This fictitious mass term, $m_v$, leads to a Majorana-tower mass $M$~\cite{majorana1937teoria}
\begin{equation}
	M = \frac {m_v}{\left[1+\left(\frac{\theta_0}{4}\right)^2 (2p+|\ell |+ 1){\cal F}(\zeta) \right]} = \frac{\hbar k_z}{c} 
\end{equation}
with the corresponding energy 
\begin{equation}
	W_0=\frac {m_v c^2}{\left[1+\left(\frac{\theta_0}{4}\right)^2 (2p+|\ell |+ 1){\cal F}(\zeta) \right]} = c \hbar k_z.
\end{equation}

Each of these beams behaves as a particle in the low energy limit that obey a Schr\"odinger/Dirac equation with a fictitious Majorana mass $M$ and a Majorana-tower OAM/fictitious mass relationship as in the 1932 paper by Majorana~\cite{Majorana:NC:1932},
\begin{equation}
	M= \frac{2 m_v}{s^\ast+\frac 12}
\end{equation}
where ${s^\ast=\theta_0^2 (2p+\abs{\ell}+1){\cal F}(\zeta)}$ is the angular momentum part of the virtual mass. In this case, the dynamics and structure of the beam is uniquely characterized in space and time through the rules of the Poincar\'e group that build
up the Majorana tower. Particles with opposite OAM values differ by chirality even if they are propagating with the same group velocity.

Furthermore, through the projection effect along the axis of propagation, $z$, one can resize and shape the beam at will to obtain different apparent sub-luminal velocities for different values of OAM and spatially buffer a set of information in time.


\section{Conclusions}
The popular (sub)class of OAM beams, obtained by imposing OAM onto a finite-sized Gaussian beam through a phase modulating device such as a spiral phase plate or a fork hologram, presents clear relationships between the beam divergence and the OAM value. This is supposed to give rise to subluminal group velocities, in particular in vacuum. We found that this relationship can be
cast in terms of a spectrum of quasiparticles with a virtual mass $m_v$ that follows the rules of the Majorana tower: the OAM value reduces the Majorana mass term, $M$, and also the group velocity, $v_g$.
The Majorana tower spectrum found here is bosonic; to obtain the fermionic/anyonic aspect, one has to apply coherent superpositions of OAM states with beams carrying non-integer values of OAM~\cite{berry}, or by introducing rotation operators
in the propagation to obtain half-integer values of angular momentum also in single photons~\cite{ballantine}.

This subclass of circular beams described by Hypergeometric-Gauss OAM confined beams thus behave as quasiparticles also in vacuum and can be used as an alternative method for the detection of the OAM value of the beam, without measuring the phase gradient. This method can be used as a precise time delay for optical communications as it is discrete and with a precise timing when emitted by sources with equal apertures.

\section*{Acknowledgements}
F.\,T. gratefully acknowledges the financial support from ZKM and MSC-BW. E.K. acknowledges the support of the Canada Research Chair (CRC) Program.

\end{document}